\documentclass[prb,twocolumn]{revtex4}
\usepackage{amsmath}
\usepackage{amsfonts}
\usepackage{graphicx}
\usepackage{bbm}
\usepackage{longtable}
\begin{document}

\title{Unsharp measurements and joint measurability} 
\author{H. S. Karthik}\email{karthik@rri.res.in}\affiliation{Raman Research Institute, Bangalore 560 080, India} 
\author{A. R. Usha Devi}\email{arutth@rediffmail.com} 
\affiliation{Department of Physics, Bangalore University, 
Bangalore-560 056, India}
\affiliation{Inspire Institute Inc., Alexandria, Virginia, 22303, USA.}
\author{A. K. Rajagopal} \email{attipat.rajagopal@gmail.com}
\affiliation{Inspire Institute Inc., Alexandria, Virginia, 22303, USA.}
\affiliation{Institute of Mathematical Sciences, C.I.T. Campus, Taramani, Chennai, 600113, India}
\affiliation{Harish-Chandra Research Institute, Chhatnag Road, Jhunsi, Allahabad 211 019, India.}
\date{\today}
\begin{abstract}
We give an overview of joint unsharp measurements of non-commuting observables using positive operator valued measures (POVMs). We exemplify the role played by  joint measurability  of POVMs in entropic uncertainty relation for Alice's pair of non-commuting observables  in the presence of  Bob's entangled quantum memory. We show that Bob  should record the outcomes of incompatible (non-jointly measurable) POVMs in his quantum memory so as to beat the entropic uncertainty bound. In other words, in addition to the presence of entangled Alice-Bob state, implementing incompatible POVMs at Bob's end is necessary to beat the uncertainty bound and hence, predict the outcomes of non-commuting observables with improved precision.  We also explore the implications of joint measurability to {\em validate} a moment matrix  constructed from average pairwise  correlations of three dichotomic non-commuting qubit observables. We prove that a classically acceptable moment matrix -- which ascertains  the existence of a legitimate joint probability distribution for the outcomes of all the three dichotomic observables -- could be realized if and only if compatible POVMs are employed. 
\end{abstract}

\keywords{unsharp measurements, incompatibility, positive operator valued measures}

\maketitle

\section{Introduction}

In the classical perspective, all physical observables can  be measured jointly. In contrast, quantum theory places restrictions on the precision with which non-commuting observables can be measured. In the conventional quantum framework, {\em sharp} measurement of an observable is done through the corresponding  spectral projection valued (PV) operators. Joint {\em sharp} values can only be assigned to 
commuting set of observables, when one  restricts to  PV measurements. More specifically, joint measurability or compatibility of observables is entirely linked with their commutativity, if {\em sharp} PV measurements are employed. However, with the introduction of  generalized measurements -- formulated in terms of positive operator valued measures (POVMs)  -- it has been possible to refine the notion of joint measurability of non-commuting observables~\cite{Ludwig64, BuschLahti84, Busch86, BuschLahti95}. While {\em sharp} values for non-commuting observables can not be assigned jointly via their PV measurements, their {\em unsharp} joint values could be perceived when  {\em compatible}   POVMs are employed. Active research efforts are dedicated to investigate operationally significant criteria  of joint measurability (compatibility) of two or more POVMs and also to develop a resource theory of  measurement incompatibility~\cite{Barnett05,Son05,Stano08,Wolf09,Yu10,LSW11,Reeb13,Kunjwal14,Brunner14,Guhne14, HSKJOSAB15, Pusey15, Heinosari15}. In this paper, we review the notion of compatible POVMs in the generalized measurement setting. We then illustrate two different physical situations, where  incompatible unsharp measurements are crucial to bring forth  non-classical features. First, we identify the significance of  incompatible POVMs to predict the  outcomes of a pair of  non-commuting observables with enhanced precision --  by focusing on the entropic uncertainty relation in the presence of quantum entangled 
memory~\cite{Berta10}. Beating the entropic uncertainty bound relies on  both entanglement and measurement incompatibility as necessary quantum resources~\cite{HSKPRA15}. Next, we explore when a moment matrix, constructed from the pairwise correlation outcomes of  joint unsharp measurements of three dichotomic non-commuting qubit observables, admits a classical joint probability distribution~\cite{HSKDMQM14}. We find that the optimal value of the unsharpness parameter, below which the moment matrix is positive (and hence, admits a joint probability distribution for the fuzzy measurement outcomes of all the three qubit observables), matches identically with the compatibility of the observables.      

The paper is organized as follows. In Sec.~II we begin with an overview of generalized measurements comprised of POVMs. We discuss the notion of joint measurability of two or more POVMs and bring out that joint measurability of POVMs is much broader than commutativity.  We explore, in Sec.~III, the implications of joint measurability on entropic uncertainty relation for Alice's  pair of non-commuting observables, in the presence of Bob's quantum memory. We show that  when Bob is  restricted to employ  only jointly measurable POVMs, it is not possible to achieve enhanced precision for predicting  Alice's measurement outcomes, even if  entangled state is shared between them. Further, in Sec. IV, we explore the role of joint measurability of three dichotomic Pauli qubit observables $\vec{\sigma}\cdot\hat{n}_k;\ k=1,2,3$ with $\hat{n}_1\cdot\hat{n}_2=\hat{n}_2\cdot\hat{n}_3=\hat{n}_1\cdot\hat{n}_3=-1/2$,  on the positivity of the moment matrix -- which is constructed from the average pairwise correlations of the  outcomes of the dichotomic observables arising in their sequential unsharp-sharp measurements. We find that the corresponding moment matrix is positive if and only if the unsharpness parameter lies in the joint measurability range of the observables. Section V is devoted to a summary of our results.             

\section{Compatible POVMs}

In the conventional  framework, quantum measurements are described in terms of  projection operators of the corresonding self-adjoint observables. And joint measurability of two commuting observables means that  one can produce the results for both the observables by performing only {\em one} PV measurement. A necessary requirement for joint measurability is that there exists  a joint probability distribution for the measurement outcomes of a set of  compatible observables, such that it yields correct marginal probability distributions for the outcomes of all the subsets of observables.  Introduction of  POVMs in 1960's by Ludwig~\cite{Ludwig64}  and  subsequent investigations on their applicability~\cite{BuschLahti84}, led to a conceptually sound and mathematically rigorous generalization of measurement theory. The usual PV measurements constitute a special case of generalized measurements. Generalized observables are represented by POVMs, which are termed as {\em unsharp} in contrast to their {\em sharp} PV counterparts.  Commutativity of POVMs has a restricted meaning than their compatibility in the framework of generalized measurements.

Mathematically, POVM  is a set $\mathbbm{E}=\{E(x)\}$ comprising of positive self-adjoint operators $0\leq E(x)\leq 1$, called {\em effects}, satisfying  $\sum_x E(x)=\mathbbm{1}$;  $x$ denotes the outcomes of measurement and $\mathbbm{1}$ is the identity operator. The notion of a POVM $\mathbbm{E}$ to be a generalized observable provides a physical representation for any possible events (effects $E(x)$) to occur as outcomes $x$ in a measurement process.    

When a quantum system is prepared in the state $\rho$, measurement of the observable $\mathbbm{E}$ gives rise to generalized L$\ddot{\rm u}$der's transformation of the state i.e., 
\begin{equation}
\rho\longmapsto \sum_x\, \sqrt{E(x)}\, \rho\, \sqrt{E(x)}  
\end{equation}
and an outcome  $x$ occurs with probability $p(x)={\rm Tr}[\rho\, E(x)]$. The expectation value of the observable is given by 
\begin{equation} 
\langle \mathbbm{E}\rangle = \sum_x x\, {\rm Tr}[\rho\, E(x)]=\sum_x x\, p(x).  
\end{equation}
The usual scenario of PV measurements is recovered as a special case, when $\{E(x)\}$ forms a set of complete, orthogonal projectors.  

A finite collection of POVMs  $\mathbbm{E}_1, \mathbbm{E}_2, \ldots, \mathbbm{E}_n$ is said to be  jointly measurable (or compatible), if there exists a {\em grand} POVM $\mathbbm{G}=\{G(\lambda);\ 0\leq G(\lambda)\leq \mathbbm{1},\, \sum_\lambda\, G(\lambda)=\mathbbm{1}\}$ from which the observables $\mathbbm{E}_i$ can be obtained  by post-processing as follows. Suppose a measurement of the global POVM $\mathbbm{G}$ is carried out in a state $\rho$ and the probability of obtaining the outcome $\lambda$ is denoted by $p(\lambda)={\rm Tr} [\rho\, G(\lambda)]$. If the effects $E_i(x_i)$ constituting the  POVM  $\mathbbm{E}_i$ can be obtained  as {\em marginals} of the {\em grand} POVM $\mathbbm{G}=\{\,G(\lambda), \lambda\equiv\{x_1,x_2,\ldots\},$ (where  $\lambda$  corresponds to a collective index $\{x_1,x_2,\ldots\}$)  i.e., if there exists a grand POVM $\mathbbm{G}$ such that~\cite{Stano08}  
\begin{eqnarray} 
\label{pp} 
E_1(x_1)&=&\sum_{x_2,x_3,\ldots}\, G(x_1,x_2,\ldots, x_n) \nonumber \\ 
E_2(x_2)&=&\sum_{x_1,x_3,\ldots}\, G(x_1,x_2,\ldots, x_n) \nonumber \\ 
\vdots && \nonumber \\ 
E_n(x_n)&=&\sum_{x_1,x_3,\ldots}\, G(x_1,x_2,\ldots, x_n),
\end{eqnarray}
 the POVMs  $\mathbbm{E}_1, \mathbbm{E}_2, \ldots, \mathbbm{E}_n$ are said to be jointly measurable. Thus, a collection of compatible POVMs $\mathbbm{E}_1, \mathbbm{E}_2, \ldots, \mathbbm{E}_n$ is obtained from a global POVM $\mathbbm{G}$ via post processing of the form (\ref{pp}).   We   emphasize once again that compatibility of POVMs does not require their commutativity,  but it demands the existence of a global POVM.  
            
As an example, consider  Pauli spin observables $\sigma_x,\, \sigma_z$  of a qubit. Sharp measurements of the  observables  
$\sigma_x=\displaystyle\sum_{x=\pm 1}\, x\, \Pi_{\sigma_x}(x)  $ and $\sigma_z=\displaystyle\sum_{z=\pm 1}\, z\, \Pi_{\sigma_z}(z)  $ is performed using the two outcome projection operators
\begin{eqnarray}
\label{sharpxy}
 \Pi_{\sigma_x}(x)&=& \frac{1}{2}\left(\mathbbm{1}+ x\, \sigma_x \right),\nonumber \\
 \Pi_{\sigma_z}(z)&=& \frac{1}{2}\left(\mathbbm{1}+  z\, \sigma_z\right).  
 \end{eqnarray}
The observables $\sigma_x$ and $\sigma_z$ are non-commuting and hence can not be measured jointly using PV measurements.    However, it is possible to consider joint fuzzy measurements of $\sigma_x, \, \sigma_z$ in terms of their POVM counterparts,  which are constructed  by adding uniform white noise to the PV operators of  (\ref{sharpxy}). One then obtains binary POVMs $\mathbbm{E}_{\sigma_x}=\{E_{\sigma_x}(x); \, x=\pm 1\}$, $\mathbbm{E}_{\sigma_z}=\{E_{\sigma_z}(z); \, z=\pm 1\}$, where 
\begin{eqnarray}
\label{unsharpxy}
 E_{\sigma_x}(x)&=&  \eta\ \Pi_{\sigma_x}(x) + (1-\eta)\, \frac{\mathbbm{1}}{2} \nonumber \\ 
& =&\frac{1}{2}\left(\mathbbm{1}+ \eta\, x\, \sigma_x \right) \nonumber \\
E_{\sigma_z}(z)&=&  \eta\ \Pi_{\sigma_z}(z) +(1-\eta)\, \frac{\mathbbm{1}}{2} \nonumber \\ 
&=&\frac{1}{2}\left(\mathbbm{1}+ \eta\, z\, \sigma_z \right) 
 \end{eqnarray} 
where $0\leq \eta\leq 1$ denotes the unsharpness parameter. It may  be noted that when $\eta=1$, the fuzzy POVMs $\mathbbm{E}_{\sigma_x}=\{E_{\sigma_x}(x)\}, \mathbbm{E}_{\sigma_z}=\{E_{\sigma_z}(z)\}$ reduce to their corresponding {\em sharp} PV versions $\{\Pi_{\sigma_x}(x)\}, \{\Pi_{\sigma_z}(z)\}$.
 
The binary POVMs  $\mathbbm{E}_{\sigma_x},\ \mathbbm{E}_{\sigma_z}$ are compatible if there exists a four element grand POVM 
$\mathbbm{G}=\{G(x,z);\ x,z=\pm 1\}$ satisfying 
\begin{widetext}
\begin{eqnarray}
\label{cond}
\sum_{z=\pm 1}\, G(x,z)&=& G(x,1)+G(x,-1)=E_{\sigma_x}(x)\nonumber \\ 
\sum_{x=\pm 1}\, G(x,z)&=& G(1,z)+G(-1,z)=E_{\sigma_z}(z)\nonumber \\ 
\sum_{x,z=\pm 1}\, G(x,z)&=&\mathbbm{1}, \ \ \ \ \  G(x,z)\geq 0.   
\end{eqnarray} 
\end{widetext}
It has been shown~\cite{Busch86, Stano08} that the POVMs $\mathbbm{E}_{\sigma_x},\ \mathbbm{E}_{\sigma_z}$ are compatible in the range $0\leq \eta\leq 1/\sqrt{2}$  of the unsharpness parameter, as it is possible to construct a global POVM $\mathbbm{G}$ comprising of the effects  
\begin{equation}
G(x,z)=\frac{1}{4}\left(\mathbbm{1}+\eta\, x\, \sigma_x + \eta\, z\, \sigma_z\right),\  0\leq \eta\leq 1/\sqrt{2}
\end{equation}
satisfying the required conditions (\ref{cond}).  Measurement of a {\em single} generalized observable (POVM) $\mathbbm{G}$ enables one to produce the results of measurement of both the  POVMs $\mathbbm{E}_{\sigma_x}$ and $\mathbbm{E}_{\sigma_z}$, when they are compatible. And, as a consequence, joint measurability of POVMs $\mathbbm{E}_{\sigma_x}$, $\mathbbm{E}_{\sigma_y}$  ensures the existence of  a joint probability distribution $p(x, z)={\rm Tr}[\rho\, G(x,z)]$ obeying $p(x)=\displaystyle\sum_z p(x, z)={\rm Tr}[\rho\, \displaystyle\sum_z\, G(x,z)]={\rm Tr}[\rho\, E_{\sigma_x}(x)]$,  $p(z)=\displaystyle\sum_x p(x, z)={\rm Tr}[\rho\, \sum_x\, G(x,z)]={\rm Tr}[\rho\, E_{\sigma_z}(z)]$, over the measurement outcomes $x,\, z$ of the unsharp POVMs $\mathbbm{E}_{\sigma_x}$, $\mathbbm{E}_{\sigma_z}$  in any arbitrary quantum state $\rho$.   

Triple-wise joint  measurements of all the three Pauli  observables $\sigma_x,\ \sigma_y$ and $\sigma_z$ 
 can be envisaged by considering  the fuzzy binary outcome POVMs  $\mathbbm{E}_{\sigma_x}=\left\{E_{\sigma_x}(x)=\frac{1}{2}\left(\mathbbm{1}+ \eta\, x\, \sigma_x \right);\, x=\pm 1\right\},$ 
 $\mathbbm{E}_{\sigma_y}=\left\{E_{\sigma_y}(y)=\frac{1}{2}\left(\mathbbm{1}+ \eta\, y\, \sigma_y \right);\, y=\pm 1\right\}$, $\mathbbm{E}_{\sigma_z}=\left\{E_{\sigma_z}(z)~=~\frac{1}{2}\left(\mathbbm{1}+ \eta\, z\, {\sigma_z} \right);\, z=\pm 1\right\}$ in the range $0\leq \eta\leq 1/\sqrt{3}$ of the unsharpness parameter~\cite{Stano08, LSW11}. Further, it has also been shown~\cite{LSW11} that the noisy versions 
 $\mathbbm{E}_{\vec{\sigma}\cdot \hat{n}_k}=\left\{E_{\vec{\sigma}\cdot \hat{n}_k}(x_k=\pm 1)=\frac{1}{2}\left(\mathbbm{1}+ \eta\, x_k\,  \, \vec{\sigma}\cdot \hat{n}_k \right)\right\}$ of the qubit spin, oriented along the unit vectors $\hat{n}_k,\ k=1,2,3$, which  are equally separated in a plane (i.e., separated by an angle $120^\circ$),  are pairwise jointly measurable if the unsharpness $\eta\leq \sqrt{3}-1$, but are triple-wise jointly measurable when $\eta\leq 2/3$. 

These examples bring forth the possibility of  quantum measurements of three observables that can be implemented jointly pairwise -- but not triplewise -- in a two dimensional Hilbert space, which could not be realized within the PV measurement framework.  This identification led towards an extension of the notion of Kochen-Specker contextuality~\cite{KS67} recently and  a generalized non-contextuality inequality~\cite{LSW11, Kunjwal14_2} is shown to be  violated  in a two dimensional Hilbert space~\cite{note2d}, if a set of three dichotomic POVMs, which have  pairwise joint measurability  --  but no triplewise joint measurability -- are employed. Moreover, it has been recognized~\cite{Busch86, Barnett05, Son05, Stano08, Wolf09, Brunner14, Guhne14} that if Bob is restricted to employ only local compatible  POVMs on his system, irrespective of Alice's measurements and of the bipartite state shared between them, a local classical probabilistic model could be realized. And hence,  it is not possible to witness non-local quantum features like  steering~\cite{Sch35, Steering} (the ability to non-locally alter the states of one part of a composite system by performing measurements on another spatially separated part) and the violation of Bell inequality~\cite{Bell64}.  An intrinsic connection between non-local steering and incompatible measurements has been independently established by  Quintino et. al.~\cite{Brunner14} and Uola et. al.~\cite{Guhne14}, who proved that a set of  POVMs is not jointly measurable if and only if it is useful to demonstrate quantum steering. In addition to bringing out the fact that measurement compatibility  is not synonymous with commutativity of the observables, these research efforts highlight the quantum resource nature of incompatible measurements. There are ongoing investigations recently, which focus towards developing a resource theory of measurement incompatibility~\cite{Pusey15, Heinosari15}.

\section{Beating entropic uncertainty bound using incompatible measurements}

In this section, we investigate  entropic uncertainty relation for Alice's pair of non-commuting observables, in the presence of an entangled quantum memory at Bob's end, when Bob is restricted to measure only compatible POVMs~\cite{HSKPRA15}.     

The Shannon entropies $H(\mathbbm{E}_X)=-\sum_x \, p(x) \log_2 p(x)$, 
$H(\mathbbm{E}_Z)=-\sum_z \, p(z) \log_2 p(z)$,  associated with the probabilities $p(x)={\rm Tr}\,[\rho\, E_{X}(x)]$, $p(z)={\rm Tr}\,[\rho\, E_{\mathbbm{E}_Z}(z)]$ of  measurement outcomes $x,\, z$ of a pair of POVMs $\mathbbm{E}_X\equiv\{E_{X}(x)\vert\, 0\leq E_X(x)\leq \mathbbm{1};\ \sum_x\, E_{X}(x)=\mathbbm{1}\} ,\ \mathbbm{E}_Z\equiv\{E_{Z}(z)\vert\, 0\leq E_{Z}(z)\leq \mathbbm{1};\ \sum_z\, E_{Z}(z)=\mathbbm{1}\}$, offer a more general framework   to quantify  uncertainties in predicting the measurement outcomes of two  observables $\mathbbm{E}_X, \mathbbm{E}_Z$ in a given quantum state $\rho$.  

The uncertainties of the measurement outcomes of  $\mathbbm{E}_X$ and $\mathbbm{E}_Z$ in a  quantum state  of finite dimension $d$ reveal a trade-off, which is expressed in terms of the entropic uncertainty relation~\cite{MU88, KP02}: 
\begin{equation} 
\label{kpbound}
H(\mathbbm{E}_X)+H(\mathbbm{E}_Z)\geq - 2\log_2\, {\cal C}(\mathbbm{E}_X,\mathbbm{E}_Z),
\end{equation}
where ${\cal C}(\mathbbm{E}_X,\mathbbm{E}_Z)={\rm max}_{x,z}\  \vert\vert \sqrt{E_{X}(x)}\, \sqrt{E_{Z}(z)}\vert\vert$. (Here, $\vert\vert A\vert\vert={\rm Tr}[\sqrt{A^\dag\, A}]$).

A generalized version of the entropic uncertainty relation, for Alice's pair of observables $\mathbbm{E}_X,\mathbbm{E}_Z$, when assisted by Bob's quantum memory,  led to a  refinement of the   uncertainty bound  (\ref{kpbound}) and brought out  that the outcomes  of  non-commuting observables could be predicted more precisely with the help of an entangled state shared between Alice-Bob~\cite{Berta10}. 

The entropic uncertainty relation in the presence of quantum memory is better introduced in terms of a game:~\cite{Berta10}: to begin with, two players Alice and Bob decide to measure a pair of observables $\mathbbm{E}_X$ and $\mathbbm{E}_Z$. Bob prepares a quantum state of his choice and sends it to Alice. Alice  measures $\mathbbm{E}_X$ or $\mathbbm{E}_Z$ randomly and communicates only her choice of measurements (not the outcomes of her measurement) to Bob. To win the game, Bob should  prepare a suitable quantum state  such that he is able to predict Alice's measurement outcomes  in every experimental run  with as much precision as possible.  In other words,  Bob's task is to minimize the uncertainties in the measurements of a pair of chosen observables $\mathbbm{E}_X, \ \mathbbm{E}_Z$, by appropriate state preparation and measurements at his end.  The uncertainties of $\mathbbm{E}_X, \ \mathbbm{E}_Z$ are bounded as in (\ref{kpbound}), if Bob can access only classical information. On the other hand, when Bob prepares an entangled state and sends one part of the state to Alice, retaining the other part, he can beat the uncertainty bound of (\ref{kpbound}).  

The entropic uncertainty relation for the observables $\mathbbm{E}_X, \mathbbm{E}_Z$, measured on Alice's subsystem of the entangled state  $\rho_{AB}$,  is given by~\cite{Berta10}  
\begin{equation} 
\label{bertabound}
S(\mathbbm{E}_X\vert B)+ S(\mathbbm{E}_Z\vert B)\geq -2\log_2 {\cal C}(\mathbbm{E}_X,\mathbbm{E}_Z) +S(A\vert B),
\end{equation}
where 
\begin{eqnarray}
S(\mathbbm{E}_X\vert B)&=&S\left(\rho^{(\mathbbm{E}_X)}_{AB}\right)-S(\rho_B)\nonumber \\    
S(\mathbbm{E}_Z\vert B)&=&S\left(\rho^{(\mathbbm{E}_Z)}_{AB}\right)-S(\rho_B) 
\end{eqnarray}
 are the conditional von Neumann entropies of the post measured states $\rho^{(\mathbbm{E}_X)}_{AB}$, $\rho^{(\mathbbm{E}_Z)}_{AB}$, which are obtained after Alice measures  $\mathbbm{E}_X,\, \mathbbm{E}_Z$  on her system and stores the outcomes $x,\ z$  in an orthonormal basis $\{\vert x\rangle\}$ ($\{\vert z\rangle\}$ respectively:
\begin{eqnarray}
\rho^{(\mathbbm{E}_X)}_{AB}&=&\sum_x\, \vert x\rangle\langle x\vert \otimes \rho^{(x)}_B \nonumber \\
\rho^{(\mathbbm{E}_Z)}_{AB}&=&\sum_z\, \vert z\rangle\langle z\vert \otimes \rho^{(z)}_B.
\end{eqnarray} 
Here, $\rho^{(x)}_B={\rm Tr}_A[\rho_{AB} (E_{X}(x) \otimes \mathbbm{1}_B)]$ and    $\rho^{(z)}_B={\rm Tr}_A[\rho_{AB} (E_{\mathbbm{Z}}(z) \otimes \mathbbm{1}_B)]$.  The probabilities of measurement outcomes $x$, $z$ are given by $p(x)={\rm Tr}[\rho^{(x)}_B]={\rm Tr}[\rho_{AB}\, (E_{X}(x) \otimes \mathbbm{1}_B)]$, \ \  $p(z)={\rm Tr}[\rho^{(z)}_B]=
{\rm Tr}[\rho_{AB}\, (E_{Z}(z) \otimes \mathbbm{1}_B)]$;  $S(A\vert B)=S(\rho_{AB})-S(\rho_B)$ denotes the conditional  von Neumann entropy of the state $\rho_{AB}$; the von Neumann entropy $S(\rho)$ of the quantum state $\rho$ is given by $S(\rho)=-{\rm Tr}[\rho\,\log_2\rho]$. 

It may be noted that when the state $\rho_{AB}$ shared between Alice and Bob  is maximally entangled, the second term on the right hand side of (\ref{bertabound}) takes the value $S(A\vert B)=-\log_2 d$. And, as the first term $- 2\log_2 C(\mathbbm{E}_X,\mathbbm{E}_Z)$ can not be larger than $\log_2 d$ (the maximum value  $\log_2 d$ for $- 2\log_2 C(\mathbbm{E}_X,\mathbbm{E}_Z)$ is realized when Alice employs pairs of unbiased projective measurements~\cite{Wehner10}),  a trivial lower bound of zero can be achieved in the entropic uncertainty relation, showing that Bob can predict Alice's outcomes with certainty. In general, by sharing  an appropriate entangled state with Alice, Bob can in fact  beat the uncertainty bound of (\ref{kpbound})  with the help of suitable measurements on his part of the state. 

Let us denote $\mathbbm{E}_{X'}$ or $\mathbbm{E}_{Z'}$ as the POVMs, which Bob choses to measure at his end, after Alice announces her choice $\mathbbm{E}_X$ or $\mathbbm{E}_Z$ of observables in each experimental run. Probabilities of Alice obtaining an outcome $x$ for  $\mathbbm{E}_X$, and Bob to get an outcome  $x'$ in his measurement of  $\mathbbm{E}_{X'}$ are given by 
\begin{equation}
\label{prob}
p(x,x')={\rm Tr}[\rho_{AB}\, E_{X}(x)\otimes E_{X'}(x')].
\end{equation}   
Shannon conditional entropies of Alice's measurement outcomes $x\in X, z\in Z$  --  conditioned  by Bob's measurement outcomes $x'\in X'$, $z'\in Z'$ -- are given by
\begin{eqnarray}
\label{shanxz}
H(X \vert X')&=&-\sum_{x, x'}\, p(x, x')\, \log_2\, p(x\vert x') \nonumber \\
H(Z \vert Z')&=&-\sum_{z, z'}\, p(z, z')\, \log_2\, p(z\vert z').
\end{eqnarray}
Here, $p(x\vert x')=p(x, x')/p(x')$ denotes the conditional probability that Alice registers an outcome $x$, when Bob finds the outcome to be $x'$; $p(x')={\rm Tr}[\rho_{B}\,E_{X'}(x')]=\sum_{x}\, p(x,x')$ is the probability of Bob getting an outcome $x'$ in the measurement of $\mathbbm{E}_{X'}$. 

Recalling that measurements can never decrease entropy, i.e.,  $H(X\vert X')\geq S(\mathbbm{E}_X\vert B)$, $H(Z\vert Z')\geq S(\mathbbm{E}_Z\vert B)$,  the entropic uncertainty relation in the presence of quantum memory (\ref{bertabound}) can also be expressed in terms of Shannon conditional entropies as, 
\begin{equation} 
\label{BobM}
H(X\vert X')+H(Z\vert Z')\geq -2\log_2 {\cal C}(\mathbbm{E}_X,\mathbbm{E}_Z) + S(A\vert B). 
\end{equation}

On the other hand, it was shown~\cite{Howell13, Walborn11} that the sum of conditional Shannon entropies $H(X\vert X' )$, $H(Z\vert Z')$ are constrained to obey the following {\em entropic steering inequality},
\begin{equation} 
\label{esr} 
H(X\vert X' )+ H(Z\vert Z') \geq -2\log_2 {\cal C}(\mathbbm{E}_X,\mathbbm{E}_Z)
\end{equation}
when Bob is unable to remotely steer Alice's state by his local measurements~\cite{Steering}. As it has been shown recently~\cite{Brunner14, Guhne14}  that measurement incompatibility is necessary and sufficient to demonstrate the violation of any steering inequality, it turns out that Bob should perform incompatible measurements at his end, so as to be able to {\em beat the uncertainty bound} of (\ref{kpbound}) below the value  `$-2\log_2 {\cal C}(\mathbbm{E}_X,\mathbbm{E}_Z)$' (or  equivalently, to witness the violation of entropic steering inequality (\ref{esr})). 
   
We now proceed to illustrate, with a particular example of qubits, how Bob can beat the upper bound on entropic uncertainties with the help of an entangled state and appropriate measurements. Suppose that Alice and Bob decide to measure a pair of  qubit observables $\sigma_x$ and $\sigma_z$ initially. Bob then prepares a pure maximally entangled two-qubit state
\begin{equation} 
\vert \psi\rangle_{AB}=\frac{1}{\sqrt{2}}\, \left(\vert 0_A,\ 1_B\rangle - \vert 1_A,\ 0_B\rangle\right)
\end{equation} 
and sends one of the subsystems to Alice.  Alice performs any one of the {\em sharp} PV measurements
\begin{eqnarray}
\label{sharpxz}
\Pi_{\mathbbm{\sigma_x}}(x)&=&\frac{1}{2}\left(\mathbbm{1}+x\, \sigma_x\, \right),\ x=\pm 1,\nonumber \\
\Pi_{\mathbbm{\sigma_z}}(z)&=&\frac{1}{2}\left(\mathbbm{1}+z\, \sigma_z\,  \right),\ z=\pm 1, 
\end{eqnarray}
 randomly on her qubit  and announces her choice ($\sigma_x$ or $\sigma_z$)  to Bob.  Bob's task is to predict Alice's measurement outcomes $x$ or  $z$ by performing suitable measurements at his end. Suppose that he performs  {\em unsharp}  measurements  
\begin{eqnarray}
\label{unsharp2}
 E_{\sigma_x}(x')&=& \frac{1}{2}\left(\mathbbm{1}+ \eta\, x'\, \sigma_x \right) \nonumber \\
E_{\sigma_z}(z')&=&  \frac{1}{2}\left(\mathbbm{1}+ \eta\, z'\, \sigma_z \right) 
\end{eqnarray} 
and announces his outcomes $x'$ or $z'$ in every experimental run.   

The joint probabilities $p(x,x')$ (or $p(z,z')$) of Alice's {\em sharp} outcome $x$ (or $z$) and Bob's {\em unsharp} outcome $x'$ (or $z'$),  are obtained (see (\ref{prob})) to be, 
\begin{eqnarray}
\label{exprob}
p(x,x')&=&\langle \psi_{AB}\vert \Pi_{\sigma_x}(x) \otimes E_{\sigma_x}(x')\vert \psi_{AB}\rangle \nonumber \\  
&=& \frac{1}{4}\left(1 - \eta\, x\, x'\right)  \nonumber \\
p(z,z')&=&\langle \psi_{AB}\vert \Pi_{\sigma_z}(z) \otimes E_{\sigma_z}(z')\vert \psi_{AB}\rangle \nonumber \\
&=& \frac{1}{4}\left(1 - \eta\, z\, z'\right).  
\end{eqnarray}
While the right-hand side of  the entropic uncertainty relation (\ref{BobM}) reduces to zero in this case, the left-hand side can be simplified (by substituting (\ref{exprob}) in (\ref{shanxz}) and simplifying) to obtain,   
\begin{widetext}
\begin{eqnarray}
\label{lhs1}
H(X\vert X')+ H(Z\vert Z')&=& -\sum_{x,x'=\pm 1}\, p(x,x') \log_2 p(x\vert x') 
-\sum_{z,z'=\pm 1}\, p(z,z') \log_2 p(z\vert z')=\, 2\, H[(1+\eta)/2]
\end{eqnarray}  
\end{widetext}
where $H(p)=-p\, \log_2 p-(1-p)\, \log_2 (1-p)$ is the Shannon binary entropy $0\leq H(p)\leq 1$. Noting that the binary entropy function $H[(1+\eta)/2]=0$ only if $\eta=1$, the trivial bound zero of the uncertainty relation (\ref{BobM})  can be achieved only when Bob too performs {\em sharp} PV measurements of the observables $\sigma_x$ and $\sigma_z$ at his end in which case  Bob can predict Alice's outcomes {\em precisely}.  Reduction in the uncertainty bound, below $-2\log_2 {\cal C}(\mathbbm{E}_{\sigma_x},\mathbbm{E}_{\sigma_z})=1$, can only be realized if $H[(1+\eta)/2]\leq 0.5$ i.e., for $\eta > 0.78$.

It may be recalled that unsharp joint measurements of the observables $\sigma_x$, $\sigma_z$ (i.e., compatibility of the  POVMs   $\mathbbm{E}_{\sigma_x}, \mathbbm{E}_{\sigma_z}$) places the restriction~\cite{Busch86, Stano08}\ \  $\eta\leq 1/\sqrt{2}\approx 0.707$ on the {\em unsharpness} parameter. If Bob confines only to  the joint measurability range $0\leq \eta\leq 1/\sqrt{2}$,  the entropic steering inequality ( see (\ref{esr}))
\begin{equation}
\label{esr2}
H(X\vert X')+ H(Z\vert Z')\geq 1     
\end{equation} 
is always satisfied. Bob's inability to steer Alice's state remotely (to be able to violate the steering inequality (\ref{esr2})) translates itself into his inability to predict Alice's outcomes with enhanced precision (i.e., to beat the entropic uncertainty bound below 1), when he is restricted to employ compatible measurements --  irrespective of the fact that he shares an entangled state  with Alice.

\section{Moment matrix positivity and measurement incompatibility}
  
Foundational conflicts about the quantum-classical worldviews of nature arise due to strikingly different statistical features in the two domains. 
Pioneering investigations by Bell~\cite{Bell64}, Kochen-Specker~\cite{KS67}, Leggett-Garg~\cite{LG85} brought out the puzzling features of probabilities of measurement outcomes,  arising within the quantum framework in terms of various no-go theorems. A common underlying feature that gets highlighted in these no-go theorems is {\em the non-existence of a joint probability distribution for the measurement outcomes of all the observables in the quantum framework}~\cite{Fine82, ARU13, TSM13, HSK_AQIS13}. 

From an entirely different perspective, {\em the classical moment problem}~\cite{Tomarkin43, Akheizer65}  
addressed the issue of determining the probability distribution, given a sequence of valid statistical moments. The classical moment problem identifies 
that a given set of real numbers qualifies to be a legitimate moment sequence of a  probability distribution, if and only if the associated moment matrix is positive. In other words, {\em existence of a valid joint probability distribution}, consistent with the  set of  moments, may be put to test in terms of the moment matrix positivity~\cite{TSM13, HSK_AQIS13}. In the present context, we focus our attention  on the role of incompatibile measurements on  the positivity of the moment matrix, constructed based on the statistics of fuzzy measurements of a set of three non-commuting dichotomic observables.   

Let $X_k, k=1,2,3$ denote the dichotomic random variables with outcomes $x_k=\pm 1$. Consider a row with four elements $\xi^T=(1, x_1x_2, x_2x_3, x_1 x_3)$. The average pairwise correlations $\langle X_k\, X_l\rangle=\displaystyle\sum_{x_k,x_l=\pm 1} p(x_k,x_l)\, x_k, x_l,\ k,l=1,2,3$ of the random variables $X_k$ can be used to construct a  $4\times 4$ moment matrix $M=\langle \xi\, \xi^T\rangle$: 
\begin{equation} 
\label{mm}
M=\left(\begin{array}{cccc} 1 & \langle X_1\,X_2\rangle & \langle X_2\, X_3\rangle & \langle X_1\, X_3\rangle \\
 \langle X_1\,X_2\rangle &  1 & \langle X_1\, X_3\rangle & \langle X_2\, X_3\rangle \\ 
\langle X_2\,X_3\rangle &  \langle X_1\, X_3\rangle & 1 &  \langle X_1\, X_2\rangle \\
\langle X_1\,X_3\rangle &   \langle X_2\, X_3\rangle & \langle X_1\, X_2\rangle & 1  
\end{array}\right).  
\end{equation}
Note that in order to obtain the diagonal elements of $M$, we find that $\langle X^2_k\, X^2_l\rangle=\displaystyle\sum_{x_k,x_l=\pm 1} p(x_k,x_l)\, x^2_k\, x^2_l= \sum_{x_k,x_l=\pm 1} p(x_k,x_l)=1$; further, the off-diagonal elements are  identified as follows: We have $M_{23}=\langle X_1\, X^2_2\,X_3\rangle=\displaystyle\sum_{x_1,x_2,x_3=\pm 1}\, p(x_1, x_2, x_3)\, x_1\, x^2_2\, x_3=\displaystyle\sum_{x_1,x_3=\pm 1}p(x_1, x_3) x_1\, x_3=\langle X_1\, X_3\rangle$ and so on, which leads to the above structure (\ref{mm}) for the moment matrix $M$ involving only average pairwise correlations of the variables. 

In the construction of $M$, it is implicit that a joint probability distribution $p(x_1, x_2, x_3)$ for the statistical outcomes of three random variables $X_1,\, X_2,\, X_3$ exists, and  pairwise probabilities $p(x_1, x_2)$, $p(x_2, x_3)$, $p(x_1, x_3)$ are obtained as  marginal distributions i.e.,   
 $\displaystyle\sum_{x_3}\, p(x_1, x_2, x_3)=p(x_1, x_2),$\ $\displaystyle\sum_{x_1}\, p(x_1, x_2, x_3)=p(x_2, x_3),$ $\displaystyle\sum_{x_2}\, p(x_1, x_2, x_3)=p(x_1, x_3)$.   
 
By construction, the moment matrix $M$ is symmetric and positive definite. And thus, all the four  eigenvalues of $M$ are positive i.e.,  
\begin{eqnarray} 
1+ \langle X_1\, X_2\rangle-\langle X_2\, X_3\rangle-\langle X_1\, X_3\rangle \geq 0  \\
1- \langle X_1\, X_2\rangle+\langle X_2\, X_3\rangle-\langle X_1\, X_3\rangle \geq 0   \\
\label{3lgi}
1- \langle X_1\, X_2\rangle-\langle X_2\, X_3\rangle+\langle X_1\, X_3\rangle \geq 0 \\
1+ \langle X_1\, X_2\rangle+\langle X_2\, X_3\rangle+\langle X_1\, X_3\rangle \geq 0 
\end{eqnarray}
Note that the positivity condition (\ref{3lgi}) on one of the eigenvalue of the moment matrix directly corresponds to three term Leggett-Garg inequality~\cite{LG85, Guhne13},  
\begin{equation}
\label{lgi3term}
\langle X_1\, X_2\rangle+\langle X_2\, X_3\rangle-\langle X_1\, X_3\rangle\leq 1.
 \end{equation} 
It has been shown~\cite{LG85, Guhne13} that sequential measurements of a set of three dichotomic quantum observables, with possible outcomes $\pm 1$, violate the inequality  (\ref{lgi3term}) -- as the average quantum pairwise correlations on the left hand side of (\ref{lgi3term}) can sum up to a maximum  value $3/2$. The maximal violation of the three term Leggett-Garg inequality  can be realized from the statistics of outcomes in the sequential PV measurements of three dichotomic qubit observables $\vec\sigma\cdot \hat{n}_k$, with  the unit vectors $\hat{n}_k,\ k=1,2,3$ equally separated in a plane by an angle $120^\circ$,  in the completely mixed initial state $\rho=\mathbbm{1}/2$ of a qubit. In other words, the moment matrix constructed based on the results of sharp PV sequential measurements of  three dichotomic qubit observables $\vec\sigma\cdot \hat{n}_k$ could turn out to be  non-positive~\cite{HSK_AQIS13} -- and hence, points towards the non-existence of a joint probability distribution $p(x_1,x_2,x_3)$ for the outcomes.    

In the following,  we consider unsharp measurements of the trine axes observables $\vec\sigma\cdot \hat{n}_k$ using the POVMs 
$\mathbbm{E}_{\vec{\sigma}\cdot \hat{n}_k}=\{E_{\vec{\sigma}\cdot \hat{n}_k}(x_k=\pm 1)\}$, where the effects $E_{\vec{\sigma}\cdot \hat{n}_k}(x_k=\pm 1)$ are given by,
\begin{equation}
E_{\vec{\sigma}\cdot \hat{n}_k}(x_k=\pm 1)=\frac{1}{2}\left(\mathbbm{1}+\eta\, x_k\, \vec{\sigma}\cdot \hat{n}_k\right).  
\end{equation}
Our intention is to obtain the range of the unsharpness parameter $\eta$ for which the $4\times 4$ moment matrix -- constructed from the average pairwise correlations $\langle X_k\, X_l\rangle$ of the three dichotomic Pauli observables -- is positive. 

We consider a maximally mixed qubit state $\rho=\mathbbm{1}/2$ and perform  sequential unsharp-sharp pairwise measurements of the observables 
$\vec{\sigma}\cdot \hat{n}_k,\ k=1,2,3$. Suppose that the first unsharp measurement of $\mathbbm{E}_{\vec{\sigma}\cdot \hat{n}_k}$ gives an outcome $x_k$. The initial quantum state  transforms to  
\begin{eqnarray}
\rho\rightarrow  \rho_k &=&\sqrt{E_{\vec{\sigma}\cdot \hat{n}_k}(x_k)}\, \rho\, \sqrt{E_{\vec{\sigma}\cdot \hat{n}_k}(x_k)}/p(x_k)\nonumber \\ 
 &=&E_{\vec{\sigma}\cdot \hat{n}_k}(x_k)/p(x_k),
 \end{eqnarray}
  with probability $p(x_k)={\rm Tr}[E_{\vec{\sigma}\cdot \hat{n}_k}(x_k)]/2=1/2$. A subsequent sharp measurement of $\vec{\sigma}\cdot \hat{n}_l$, yielding an outcome $x_l$, results in the post measured state 
$\Pi_{{\vec{\sigma}\cdot \hat{n}_k}(x_l)}\, E_{\vec{\sigma}\cdot \hat{n}_k}(x_k)\, \Pi_{{\vec{\sigma}\cdot \hat{n}_k}(x_l)}/p(x_l\vert x_k)$
with the probability  $p(x_l\vert x_k)$ of obtaining the sharp outcome $x_l$ for $\vec{\sigma}\cdot \hat{n}_l$,  given that the first unsharp measurement of $E_{\vec{\sigma}\cdot \hat{n}_k}$ has resulted in the outcome $x_k$   given by 
\begin{eqnarray}
p(x_l\vert x_k)&=&{\rm Tr}[E_{\vec{\sigma}\cdot \hat{n}_k}(x_k)\, \Pi_{{\vec{\sigma}\cdot \hat{n}_k}(x_l)}]\nonumber \\
&=& \frac{1}{2}\, \left(1+\eta\, x_k\, x_l\, \hat{n}_k\cdot\hat{n}_l\right) \nonumber \\
&=& \frac{1}{2}\, \left(1-\frac{\eta}{2}\, x_k\, x_l\, \right).   
\end{eqnarray}
 The pairwise joint probabilities $p(x_k,x_l)$ of the sequential measurement are then evaluated as, 
 \begin{eqnarray}
p(x_l, x_k)&=&p(x_k)\, p(x_l\vert x_k)\nonumber \\
&=& \frac{1}{4}\, \left(1-\frac{\eta}{2}\, x_k\, x_l\right).         
 \end{eqnarray}
Using the above  joint probabilities, the average pairwise correlations $\langle X^{(\rm u)}_k X^{(\rm s)}_l\rangle, k < l=1,2,3$ of the unsharp-sharp sequential measurements are evaluated to obtain, 
\begin{eqnarray}
\langle X^{(\rm u)}_1 X^{(\rm s)}_2\rangle &=& \sum_{x_1,x_2}\, p(x_1, x_2)\, x_1\, x_2= -\eta/2 \nonumber \\      
\langle X^{(\rm u)}_2 X^{(\rm s)}_3\rangle\rangle&=&\sum_{x_2,x_3}\, p(x_2, x_3)\, x_2\, x_3 = -\eta/2 \nonumber \\
\langle X^{(\rm u)}_1 X^{(\rm s)}_3\rangle&=&\sum_{x_1,x_3}\, p(x_1, x_3)\, x_1\, x_3 = -\eta/2. 
\end{eqnarray}
The corresponding $4\times 4$ moment matrix (see (\ref{mm})) is then given by,  
\begin{equation} 
M=\left(\begin{array}{cccc} 1 & -\eta/2 & -\eta/2 & -\eta/2 \\
 -\eta/2 &  1 & -\eta/2 & -\eta/2 \\ 
-\eta/2 &  -\eta/2 & 1 &  -\eta/2 \\
-\eta/2 &   -\eta/2 & -\eta/2 & 1  
\end{array}\right).  
\end{equation}
The eigenvalues of the moment matrix are readily found to be $\lambda_1=(2+\eta)/2=\lambda_2=\lambda_3$ and $\lambda_4=(2-3\eta)/2$. Positivity of the moment matrix  implies that  $\eta\leq 2/3$, which matches exactly with the range of the unsharpness parameter over which the POVMs $\mathbbm{E}_{\vec{\sigma}\cdot\hat{n_k}}, k=1,2,3$ for the trine axes $\hat{n}_k$ are compatible~\cite{LSW11}. We thus obtain the result that moment matrix positivity  and joint measurability of the observables are synonymous. 

\section{Conclusions} 
 In the classical framework, physical observables are all compatible and they can be measured jointly. In contrast, measurement of observables, which do not commute, are declared to be incompatible in the quantum scenario. The notion of compatibility of measurements is synonymous  with commutativity of the observables only when one restricts to PV measurements. A broader notion of compatibility emerged in a generalized framework of unsharp measurements using POVMs. In the generalized measurement theory, joint measurability of  two or more POVMs does not, in general, require their commutativity, but it necessarily requires the existence of a grand POVM, measurement of which suffices to construct the results of measurments of the set of compatible (jointly measurable) POVMs. In this paper we have reviewed the notion joint measurability  of POVMs. We have also given a detailed discussion on the importance of incompatible measurements, to be employed by Bob at his end, so as to  beat the entropic uncertainty bound for a pair of non-commuting observables of Alice's spatially separated quantum system, entangled with Bob's state.  Further, we have  brought out the connection between  measurement compatibility and positivity of a moment matrix,  by considering a specific example of sequential unsharp-sharp measurements of pairs of  qubit observables $\vec{\sigma}\cdot\hat{n}_k;\ k=1,2,3$, where the three unit vectors $\hat{n}_1,\, \hat{n}_2,\, \hat{n}_3$ are equally separated in a plane, making an angle $120^\circ$. Specifically, we have shown that the moment matrix is positive if and only if the unsharpness parameter $\eta \leq 2/3$, which coincides exactly with that for the joint measurability of all the three qubit observables $\vec{\sigma}\cdot\hat{n}_k;\ k=1,2,3$. Our example indicates that positivity of the  moment matrix, existence of joint probabilities, and compatibility of POVMs are all equivalent notions.


\begin{thebibliography}{0} 


\bibitem {Ludwig64} G. Ludwig, Z. Physik, {\bf 181} 233, (1964).

\bibitem{BuschLahti84} P. Busch and P. Lahti, 
Phys. Rev. D, {\bf 29} 1634, (1984).

\bibitem{Busch86} P. Busch, 
Phys. Rev. D, {\bf 33} 2253, (1986).

\bibitem{BuschLahti95} P. Busch, M. Grabowski, and P. J. Lahti, {\em Operational Quantum Physics} (Springer-Verlag, Berlin, 1995). 


\bibitem{Barnett05} E. Andersson, S. M. Barnett, and A. Aspect, \pra {\bf 72}, 042104 (2005).

\bibitem{Son05} W. Son, E. Andersson, S. M. Barnett, and M. S. Kim, Phys. Rev. A
{\bf 72}, 052116 (2005). 

\bibitem{Stano08} T. Heinosaari, D. Reitzner, and P. Stano, Found. Phys. {\bf 38}, 1133
(2008). 

\bibitem{Wolf09} M. M. Wolf, D. Perez-Garcia, and C. Fernandez, Phys.
Rev. Lett. {\bf 103}, 230402 (2009).

\bibitem{Yu10} S. Yu, N. Liu, L. Li, and C. H. Oh, \pra {\bf 81}, 062116 (2010). 

\bibitem{LSW11} Y. C. Liang, R. W. Spekkens, and H. M. Wiseman, Phys. Rep.
{\bf 506}, 1 (2011).

\bibitem{Reeb13} D. Reeb, D. Reitzner, and M. M. Wolf, J. Phys. A: Math.
Theor. {\bf 46}, 162002 (2013).

\bibitem{Kunjwal14} R. Kunjwal, C. Heunen, and T. Fritz, 
Phys. Rev. A {\bf 89}, 052126 (2014).

\bibitem{Brunner14} M. T. Quintino, T. V{\' e}rtesi, and N. Brunner, Phys. Rev. Lett. {\bf 113}, 160402 (2014).

\bibitem{Guhne14} R. Uola, T. Moroder, and O. G{\" u}hne, Phys. Rev. Lett. {\bf 113}, 160403 (2014).


\bibitem{HSKJOSAB15} H. S. Karthik, J. Prabhu Tej, A. R. Usha Devi, and A. K. Rajagopal, J. Opt. Soc. Am. B {\bf 32}, A34 (2015).


\bibitem{Pusey15} M. Pusey, J. Opt. Soc. Am. {\bf 32}, A56 (2015). 

\bibitem{Heinosari15} T. Heinosaari, J. Kiukas, D. Reitzner, and J. Schultz, arXiv:1504.05768. 

\bibitem{Berta10} M. Berta, M. Christandl, R. Colbeck, J. M. Renes, and R. Renner, Nature Physics {\bf 6}, 659 (2010).

\bibitem{HSKPRA15} A. R. Usha Devi, Invited talk at the Discussion Meeting on Quantum Measurements (DMQM) held at Indian Institute of Science, Bangalore during 22-24 October, 2014; H. S. Karthik, A. R. Usha Devi and A. K. Rajagopal, \pra {\bf 91}, 012115 (2015).

\bibitem{HSKDMQM14} H. S. Karthik and A. R. Usha Devi, Poster presented at the Discussion Meeting on Quantum Measurements (DMQM) held at Indian Institute of Science, Bangalore during 22-24 October, 2014.   

\bibitem{KS67} S. Kochen and E. P. Specker, J. Math. Mech. 17, 59 (1967).


\bibitem{Kunjwal14_2} R. Kunjwal and S. Ghosh, Phys. Rev. A. {\bf 89}, 042118 (2014).

\bibitem{note2d}  Conventionally, quantum violation of  Kochen-Specker non-contextuality~\cite{KS67} (which employs  three pairwise compatible, but not tripplewise compatible observables) could be illustrated for  Hilbert spaces dimension $d\geq 3$, when only projective measurements were employed.  

\bibitem{Sch35} E. Schr{\" o}dinger, Proc. Camb. Phil. Soc. {\bf 31}, 555 (1935). 

\bibitem{Steering} Apart from Bell non-locality, yet another manifestation of non-locality called 'steering'  has attracted attention recently. Though the striking feature that it is possible to remotely {\em steer} the state of a  pary (Alice) through local measurements on the state of another distant party (Bob) was noticed by Erwin Schr{\" o}dinger~\cite{Sch35} as early as 1935, it was Reid (Phys. Rev. A {\bf 40}, 913 (1989)), who proposed the first experimentally testable critera of non-local steering  to show that  steering  and Einstein-Podolsky-Rosen paradox are equivalent notions of non-locality. Further,  Wiseman et.al.~( Phys. Rev. Lett. {\bf 98}, 140402) formally introduced the notion of local hidden state to investigate steering phenomena. They showed that steering constitutes a  different class of non-locality that lies between entanglement and Bell non-locality. For a comprehensive historical outline and for the modern formalism of quantum steering see  E. G. Cavalcanti, S. J. Jones, H. M. Wiseman, and M. D. Reid, Phys. Rev. A {\bf 80}, 032112 (2009).

\bibitem{Bell64}  J. S. Bell, Physics {\bf 1}, 195 (1964).


\bibitem{MU88} H. Maassen and J. B. Uffink, Phys. Rev. Lett. {\bf 60}, 1103 (1988).

\bibitem{KP02} M. Krishna and K. R. Parthasarathy, Sankhya {\bf 64}, 842 (2002).  


\bibitem{Wehner10} S. Wehner and A. Winter, New J. Phys. {\bf 12}, 025009 (2010). 

\bibitem{Howell13}  J. Schneeloch, C. J. Broadbent, S. P. Walborn, E. G. Cavalcanti, and J. C. Howell, \pra {\bf 87}, 062103 (2013).     

\bibitem{Walborn11} The first entropic criteria of steering was formulated for position and momentum by  S. P. Walborn, A. Salles, R. M. Gomes, F. Toscano, and P. H. Souto Ribeiro, Phys. Rev. Lett. 106, 130402 (2011). Entropic steering inequalities for discrete observables were developed more recently~\cite{Howell13} by Schneeloch et. al. 

\bibitem{LG85} A. J. Leggett and A. Garg, Phys. Rev. Lett. {\bf 54}, 857 (1985).

\bibitem{Fine82} A. Fine, Phys. Rev. Lett. {\bf 48}, 291 (1982); {\em ibid}., J. Math. Phys. {\em 23}, 1306 (1982).

\bibitem{ARU13} A. R. Usha Devi, H. S. Karthik,  Sudha and  A. K. Rajgaopal, \pra {\bf 87}, 052103 (2013).  
  
\bibitem{TSM13}  H. S. Karthik, H. Katiyar, A. Shukla, T. S. Mahesh, A. R. Usha Devi and A. K. Rajagopal, \pra {\bf 87}, 052118 (2013).
\bibitem{HSK_AQIS13} H. S. Karthik, A. R. Usha Devi, A. K. Rajagopal, Sudha, J. Prabhu Tej and A. Narayanan, Proceedings of 13th Asian Quantum Information Science Conference (AQIS),  38 (2013). 
\bibitem{Tomarkin43} J.A. Shohat and J.D. Tamarkin, {\em The problem of mo-
ments}, (American Mathematical Society, 1943).


\bibitem{Akheizer65} N.I. Akhiezer, {\em The Classical Moment Problem}, (Hafner
Publishing Co., New York, 1965).

\bibitem{Guhne13} C. Budroni, T. Moroder, M. Kleinmann, and O. G{\" u}hne, \prl {\bf 111}, 020403 (2013). 
 
\end{thebibliography}
\end{document}